# Accelerated expansion of the universe and chasing photons from the CMB to study the late time integrated Sachs-Wolfe effect over different redshift ranges

Syed Faisal ur Rahman (1,2), Muhammad Jawed Iqbal (1)

Corresponding Author: Faisalrahman36@hotmail.com

1) Institute of Space and Planetary Astrophysics (ISPA), University of Karachi (UoK), Karachi, Pakistan.
2) Karachi Institute of Technology and Entrepreneurship (KITE), Karachi, Pakistan.

**Abstract:** In this study, we are going to discuss the accelerated expansion of the universe and how this accelerated expansion affects the paths of photons from cosmic microwave background radiation (CMB). Then we will see how wide-field galaxy surveys along with cosmic CMB anisotropy maps can help us in studying dark energy. The cross-correlation of galaxy over/under-density maps with CMB anisotropy maps help us in measuring one of the most useful signatures of dark energy i.e. Integrated Sachs-Wolfe (ISW) effect. ISW effect explains the blue-shifting and red-shifting of CMB photons when they reach to us after passing through large scale structures and super-voids respectively. We will look into the theoretical foundations behind ISW effect and discuss how modern all sky galaxy surveys like EMU-ASKAP will be useful in studying the effect over different redshift ranges.

# INTRODUCTION

Since the discovery of the accelerating expansion of the universe in late 1990s [1, 2, 3, 4], cosmologists are working on different signatures and theories to explain the phenomenon. Type 1a supernovae, which provided the initial evidence of accelerated expansion and thus highlighted the importance of dark energy, are still being used to study the phenomenon especially when it comes to constraining the dark energy equation of state parameter.

Galaxy surveys also provide some of the most important clues in understanding dark energy, especially in combination with cosmic microwave background surveys, which can be used to calculate phenomena such as Integrated Sachs Wolfe (ISW) effect [5, 6, 7, and 8], gravitational lensing, cosmic magnification, galaxy matter power spectrum and others. These phenomena help us in constraining parameters related to dark energy. We will study ISW effect as a tool to constraint dark energy parameters. ISW effect uses spatial cross-correlation between large scale structure tracers and cosmic microwave background anisotropy maps to constraint cosmological parameters including dark energy density parameter.

## Accelerated expansion and Type 1a Supernovae

During the late 1990s, discovery of the accelerated expansion of the universe by two different teams, High Z Supernova team and Supernova Cosmology Project team, took the physics world by surprise [9, 1, 10, 3, 4, and 11]. This discovery led to the end of matter dominated universe as it believed at that time. The fact that universe is expanding at an accelerating pace required certain adjustments in the cosmological models of that time.

One big adjustment was the concept of dark energy and to represent it in a dark matter dominated universe, the concept of dark energy parameter was introduced in cosmological models.

Several models were proposed to explain the phenomenon involving cosmological constant (such as: ΛCDM) or involving more general dark energy variables (such as: wCDM) and various other models.

Supernovae type 1a, can be used to trace the expansion history of our universe by measuring their luminosity distance (DL) via redshift information, absolute magnitude (M) and apparent magnitude (m) which uses both luminosity distance and absolute magnitude and input parameters.

Apparent magnitude (m) can be written as:

$$m(C,z)=5\log[DL(C,z)]+M+25 \qquad (1)$$

Here, C is used as set of assumed cosmological parameters.

We then need to measure the distance modulus as:

$$\mu(C,z)= m(C,z) - M \qquad (2)$$

Using (1), we can re-write distance modulus as:

$$\mu(C,z)= 5\log[DL(C,z)] + 25 \qquad (3)$$

In FLRW Universe with metric written as:

$$ds^2 = -c^2 d\tau^2 = -c^2 - dt^2 + a(t)^2 \left(\frac{dr^2}{1-kr^2} + r^2 \left(d\theta^2 + \sin^2\theta \, d\phi^2\right)\right) \qquad (4)$$

Considering k=0, for a spatially flat universe, we can write luminosity distance as:

$$DL(z) = (1+z)\, \chi(z) \qquad (5)$$

Where, $\chi(z)=c\eta(z)$ is the commoving distance and $\eta(z)$ is conformal loop back time which can be calculated as:

$$\eta(z) = \int_0^z \frac{dz'}{H(z)} = \int_0^z \frac{dz'}{H0E(z')} \qquad (6)$$

Here, H0 is the Hubble constant and $E(z)=\sqrt{\Omega_\Lambda+\Omega r(1+z)^4+\Omega m(1+z)^3}$ for flat Lambda-CDM model. To test models, we usually use the chi-square ($\chi$^2) test. We look for the minimal value of $\chi$^2 against our model parameters and given data.

$$\chi^2 = \Sigma \left(\frac{\mu model-\mu i}{\sigma i}\right)^2 \qquad (7)$$

Using the minimum $\chi$^2 test we can check values for various standard deviations thresholds. This will help us draw constraints over dark energy density and other parameters in the model.

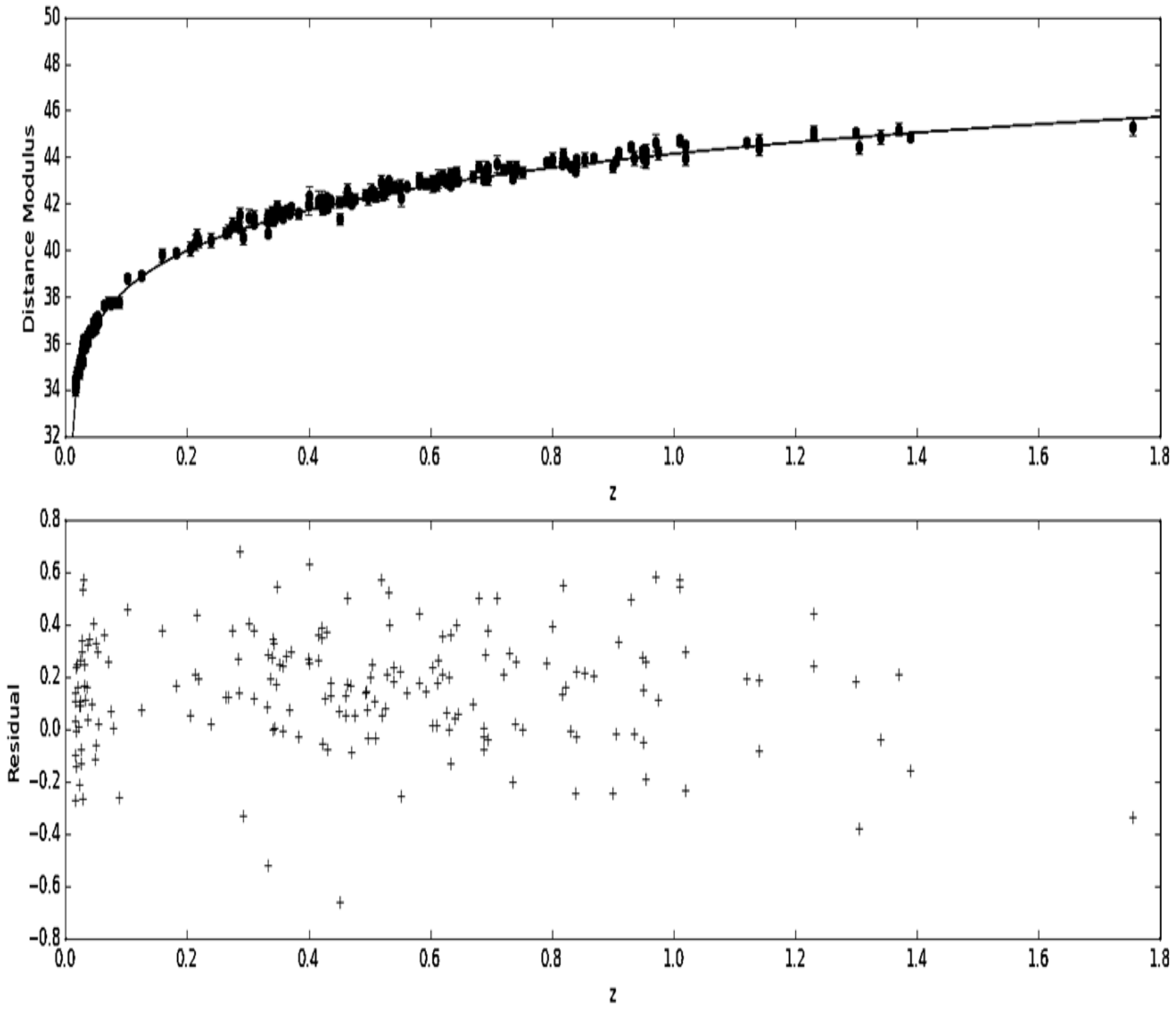

**Figure 1-Distance modulus plots for type 1a supernovae data using WMAP 9 parameters**

For figure (1), we used WMAP 9 years cosmological parameters with flat cosmological model and calculated the $\chi^2$/degrees of freedom value ≈1.5 and $\chi^2$ ≈291. The plot was produced using the data set from Davis et al. 2007[10] which used a combine dataset from Wood & Vasey et al. 2007 [12] and Riess A. et al. 2007 [4]. Type 1a supernova results may hint towards the accelerated expansion of the universe but to constraint the cosmological parameters we need to look towards something which will give us a chance to go deeper and look wider in the space for more significant results [5, 11]. In the next section, we will see how this accelerated expansion affects the path of photons from the cosmic microwave background radiation (CMB) while going through large scale structures and super-voids. The phenomenon is known as the late time Integrated Sachs-Wolfe effect [13, 14, 15, 16, 17, 18, 7, 8, 19, and 20]. We will also discuss the effect in the context of the upcoming Evolutionary Map of the Universe (EMU) survey which will be one of the main science goals of the Australian Square Kilometer Array Pathfinder (ASKAP) radio telescope array [21, 22, and 5].

## ISW Effect

Late time Integrated Sachs-Wolfe Effect or simply ISW Effect, explains the blue shifting and red shifting of the CMB photons due to the presence of large scale structures or heavy gravitational potential and supervoids respectively. ISW effect is an important tool to study dark energy which is used to account for the accelerate expansion of the universe and to understand problems related to the age of the universe and cosmic distances [23, 24, 25, 26, 27, 28, 29, 30, 31, 32, and 33].

We can write total temperature perturbation as [30, 5, and 6]:

$$\frac{\delta T(\boldsymbol{n}, \eta 0)}{T} = \frac{1}{4}\delta\gamma(\eta r) + \Phi(\eta r) + \int_{\eta r}^{\eta 0} (\Phi' - \Psi')d\eta + \boldsymbol{nv}\,(\eta r) \tag{8}$$

First two parts of the equation represent the early Sachs-Wolfe effect, third parts is the late time Integrated Sachs-Wolfe effect which is the main focus of this paper and the final part represents the Doppler effect when baryon-electron-photon medium moves with respect to the conformal Newtonian frame with the velocity v(ηr).

Before moving forward we should make few assumptions about our simplistic model. We are working in Λ CDM universe with:

1) Linear Perturbations
2) $\Phi' = 0$ for matter domination
3) $\Phi' \neq 0$ for dark energy domination universe
4) Flat, homogenous and isotropic universe
5) Ideal cosmological fluids
6) Non-relativistic matter

In eq. (8), the third part $\int_{\eta r}^{\eta 0} (\Phi' - \Psi') d\eta$ is known as Integrated Sachs-Wolfe Effect [31, 32, 34, and 35]. The other parts represent early time perturbations but the ISW part is late time. The early parts are from the photon last scattering epoch at which linearized theory is applicable. But $\int_{\eta r}^{\eta 0} (\Phi' - \Psi') d\eta$ is late ISW so linearized model is not applicable in this regime. But for large angular scales which corresponds to large spatial scales of perturbations linear assumption stills works.

The gravitational potential are constant at matter dominated universe so,

$\Phi' = \Psi' = 0.$

And since on large scales, we treat the process as linear so no integrated Sachs-Wolfe effect in the matter dominated universe. This information is particularly useful in studying the ISW effect for higher redshifts as the effect is likely to start diminishing as we go towards higher redshifts ($z > 1.5 \sim 2$).

However, if we move away from the linear regime to non-linear then the effect shows its presence and is known as Rees-Sciama effect [36].

In linearized equations, the metric with scalar perturbations is [36, 37, 38, and 39]:

$$ds^2=a^2(\eta)[(1+2\Phi)d\eta^2 - (1+2\Psi)dx^2] \quad (9)$$

The components of Einstein tensor in the conformal Newtonian gauge are [30]:

$$\delta G^0{}_0=\frac{2}{a^2}\left(-\Delta\Psi + \frac{3a'}{a}\Psi' - \frac{3a'^2}{a^2}\Phi\right) \quad (10)$$

$$\delta G^0{}_i=\frac{2}{a^2}\left(-\partial i\Psi' + \frac{a'}{a}\partial i\Phi'\right) \quad (11)$$

$$\delta G^i{}_j=\frac{1}{a^2}\partial i\partial j(\Phi+\Psi) - \frac{2}{a^2}\delta ij\left[-\Psi'' + \frac{1}{2}\Delta(\Phi+\Psi) + \frac{a'}{a}(\Phi' - 2\Psi') + \frac{2a''}{a}\Phi - \frac{a'^2}{a^2}\Phi\right] \quad (12)$$

In linearized equations in the conformal Newtonian gauge, we can take [30, 6]:

$$\Psi = -\Phi$$

Now the Einstein equations become [30]:

$$\Delta\Phi - \frac{3a'}{a}\Phi' - \frac{3a'^2}{a^2}\Phi = 4\pi G a^2 . \delta\rho tot \quad (13)$$

$$\Phi' + \frac{a'}{a}\Phi = -4\pi G a^2 . [(\rho+p)v)tot \quad (14)$$

$$\Phi'' + 3\frac{a'}{a}\Phi' + \left(\frac{2a''}{a} - \frac{a'^2}{a^2}\right)\Phi = 4\pi G a^2 . \delta p tot \quad (15)$$

Also, by using the relationship between Ψ and Φ, we can write ISW effect as [36, 40, 41, 5, and 6]:

$$\frac{\delta T(ISW)}{T} = 2\int_{\eta r}^{\eta 0} \Phi' d\eta \quad (16)$$

Equation (9) can also be written as by using the relationship between Ψ and Φ as:

$$ds^2 = ds^2 = a^2(\eta)[(1+2\Phi)d\eta^2 - (1-2\Phi)dx^2] \quad (17)$$

We can write equations (13) and (15), by using momentum representation, as [30, 6]:

$$k^2\Phi + \frac{3a'}{a}\Phi' + \frac{3a'^2}{a^2}\Phi = -4\pi G a^2 . \delta\rho \quad (18)$$

$$\Phi'' + 3\frac{a'}{a}\Phi' + \left(\frac{2a''}{a} - \frac{a'^2}{a^2}\right)\Phi = 4\pi G a^2 . \delta p \quad (19)$$

The equation of state for perturbations is:

$$\delta p = u_s^2 \delta \rho \quad (20)$$

Where, $u_s$ is the arbitrary sound speed, p is pressure and ρ is density .

For our assumed universe, we can use, for continuing dark energy domination [30]:

$$a(\eta) = \frac{-1}{H_{ti}\,\eta} = a \quad (\eta < 0)$$

Where η is conformal time and $H_{ti}$ is time independent Hubble parameter.

$$H^2_{ti} = (8\pi/3)G\rho_\Lambda$$

$$a' = 1/H\eta^2$$

$a'' = -2/H\eta^3$

Assuming that dark energy is constant in space and time i.e. $\delta\rho_\Lambda=0$ and using the relations for a, a' and a'' in equation (19). Then we get [30, 6]:

$$\Phi'' - 3\frac{\Phi'}{\eta} + \frac{3}{\eta^2}\Phi = 0 \qquad (21)$$

Giving

$$\Phi(\eta) = C_1\eta + C_2\eta^3 \qquad (22)$$

Showing $\Phi \propto \eta \propto (1/a)$ and $\Phi \propto \eta^3 \propto (1/a^3)$

Now $\Phi' = C_1 + 3C_2\eta^2$ (23)

$\Phi' \propto \eta^2\ (1/a^2)$

In standard cosmology, where the local gravitational potential is related to the matter distribution via the Poisson equation [40, 41, and 30]:

$$\nabla^2\Phi = 4\pi G a^2 \rho_m \delta_m$$

Here $\rho_m$ is matter perturbation, $\Phi$ is gravitational potential and $\delta_m$ is matter over/under-density perturbation.

We have relation:

$$\Phi(z,k) \equiv g(z)\Phi_{matter_dominated}(k)$$

Where g(z) is the suppression factor.

Also, we have:

$3H_0^2/8\pi G\rho = 1/\Omega$

Here $H_0$ is the Hubble Constant, ρ is density and Ω is the density parameter.

Using these relations and rewriting $\nabla^2\Phi$ relation in Fourier space will make $\nabla^2$ $to - k^2$.

$$\Phi(k,\eta) = -\frac{3}{2}\Omega m \left(\frac{Ho}{ck}\right)^2 g(\eta)\delta(k)$$

Here g(η )=D(η )/a(η ) is the linear growth suppression factor, $\Omega m$ is the matter density parameter and δ (k) is the matter over/under-density field [40].

Temperature fluctuations provided in eq. (8) can be written in terms of spherical harmonics as:

$$\Delta\mathrm{T}(\theta,\Phi) = \sum_{l=2}^{l=\infty}\sum_{m=-l}^{m=+l} almYlm(\theta,\Phi)$$

Here ΔT is temperature fluctuation, Ylm values are spherical harmonics and alm coefficient represents deviations from the mean values in a given map. In a smooth map alm will be zero. A more useful way of measuring these fluctuations is in terms of the angular power spectrum, which can be written as [42]:

$$Cl = \langle |a_{lm}|^2 \rangle$$

Theoretically, these Cl values can be calculated as:

$Cl = \frac{2}{\pi}\int k^2 dk\, P(k)[Wl^t(k)]^2$

Window function for temperature anisotropy due to gravitational potential can be written as [5,40]:

$W_l^t(k) = 3\Omega m \left(\frac{Ho}{ck}\right)^2 \int \frac{\partial \Phi}{\partial z} jl[ck\eta(z)]dz$

Using the Limber approximation [43, 44], we can re-write Cl as:

$$Cl^{ISW} = T^2cmb\left(\frac{3Ho^2\Omega m}{c^2}\right)^2 \frac{1}{\left(l+\frac{1}{2}\right)^4}\int dz\frac{\chi(z)^2H(z)}{c}\left(\frac{d(1+z)D(z)}{dz}\right)^2 P((l\ +0.5)/\chi(z)). \qquad (24)$$

The total temperature fluctuations in the CMB [18] maps are much larger as compared to the late time Integrated Sachs-Wolfe effect alone, as shown in the figure(2).

However, using theoretical plots we can get some idea about the strength of the expected signal at various redshift ranges figures (3, 4, 5, 6 & 7).

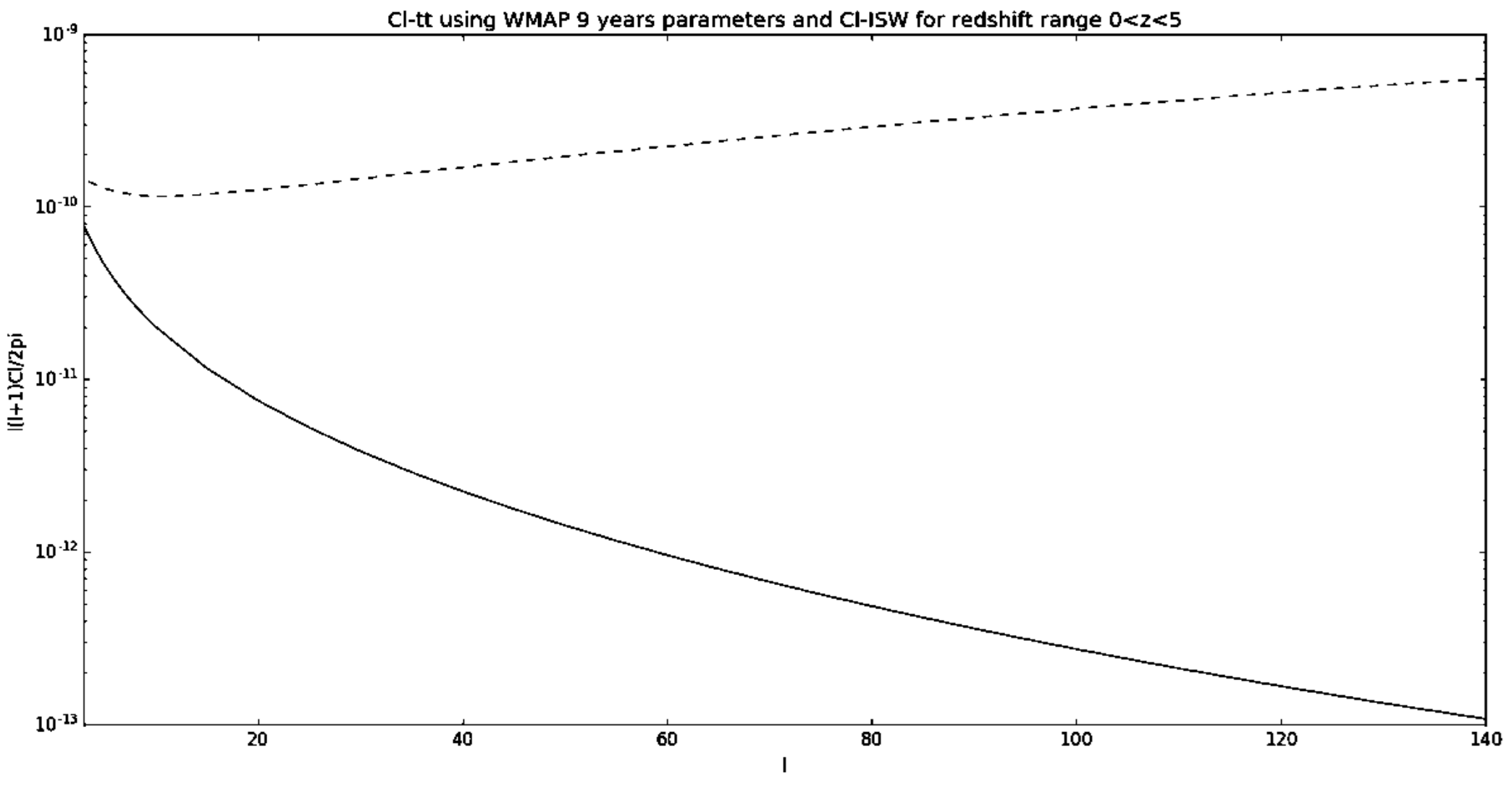

**Figure 2- Comparison of whole CMB anisotropy angular power spectrum with only ISW signal. We can clearly see that the ISW signal is relatively insignificant as compared to the whole CMB signal consequently making it very unlikely to be detected directly.**

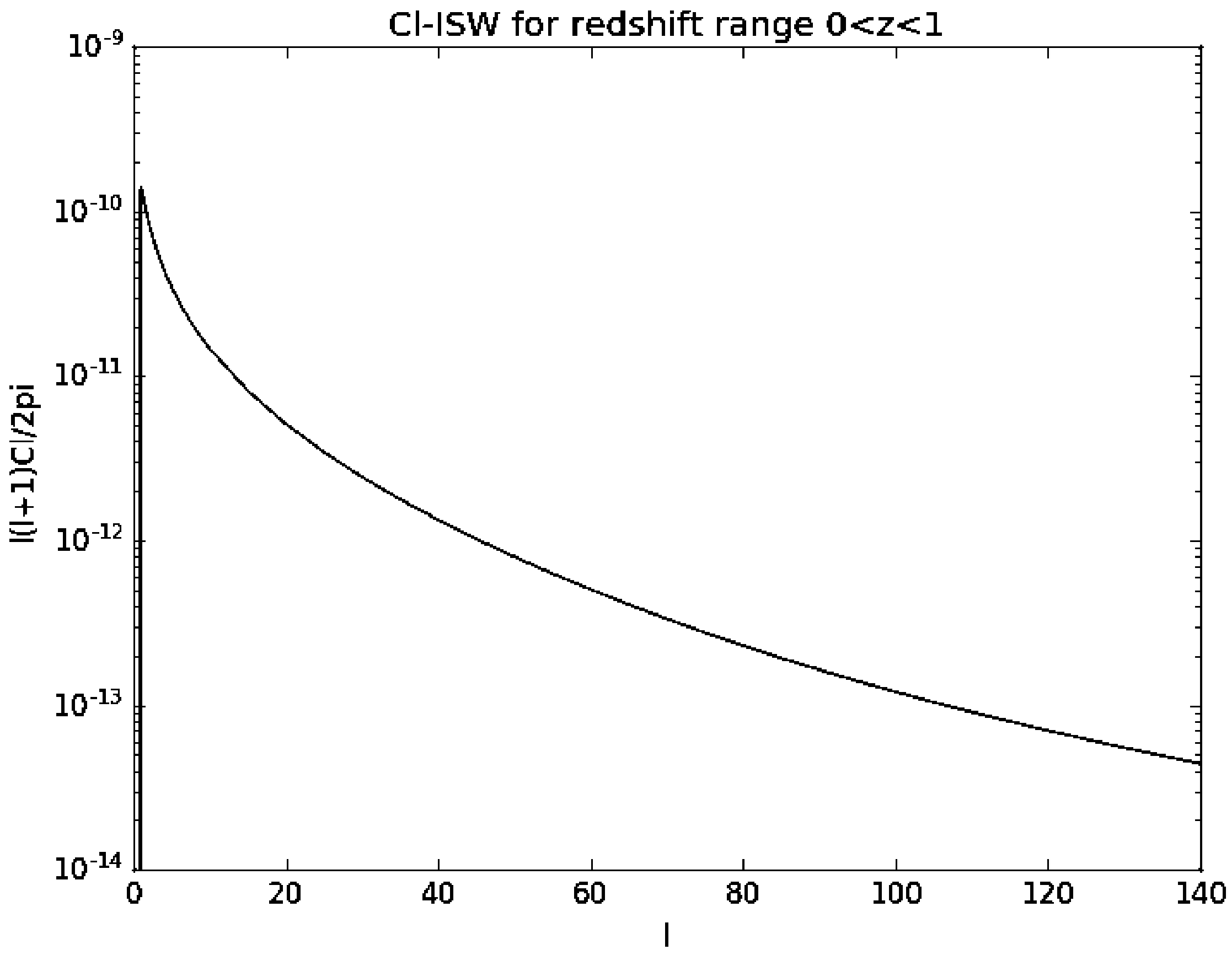

Figure 3-Theoretical ISW angular power spectrum between redshifts 0 and 1 using WMAP 9 years parameters.

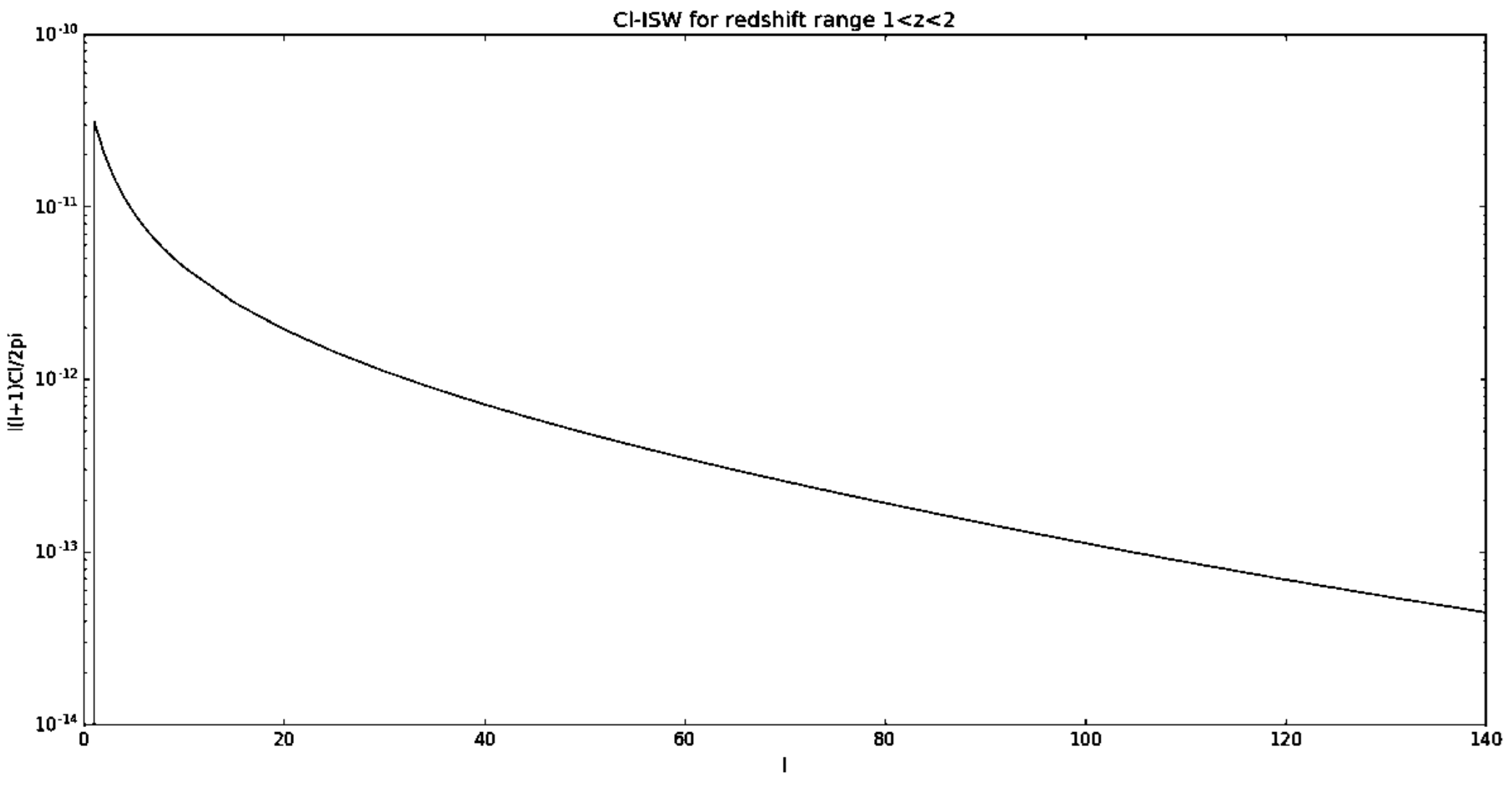

**Figure 4-Theoretical ISW angular power spectrum between redshifts 1 and 2 using WMAP 9 years parameters.**

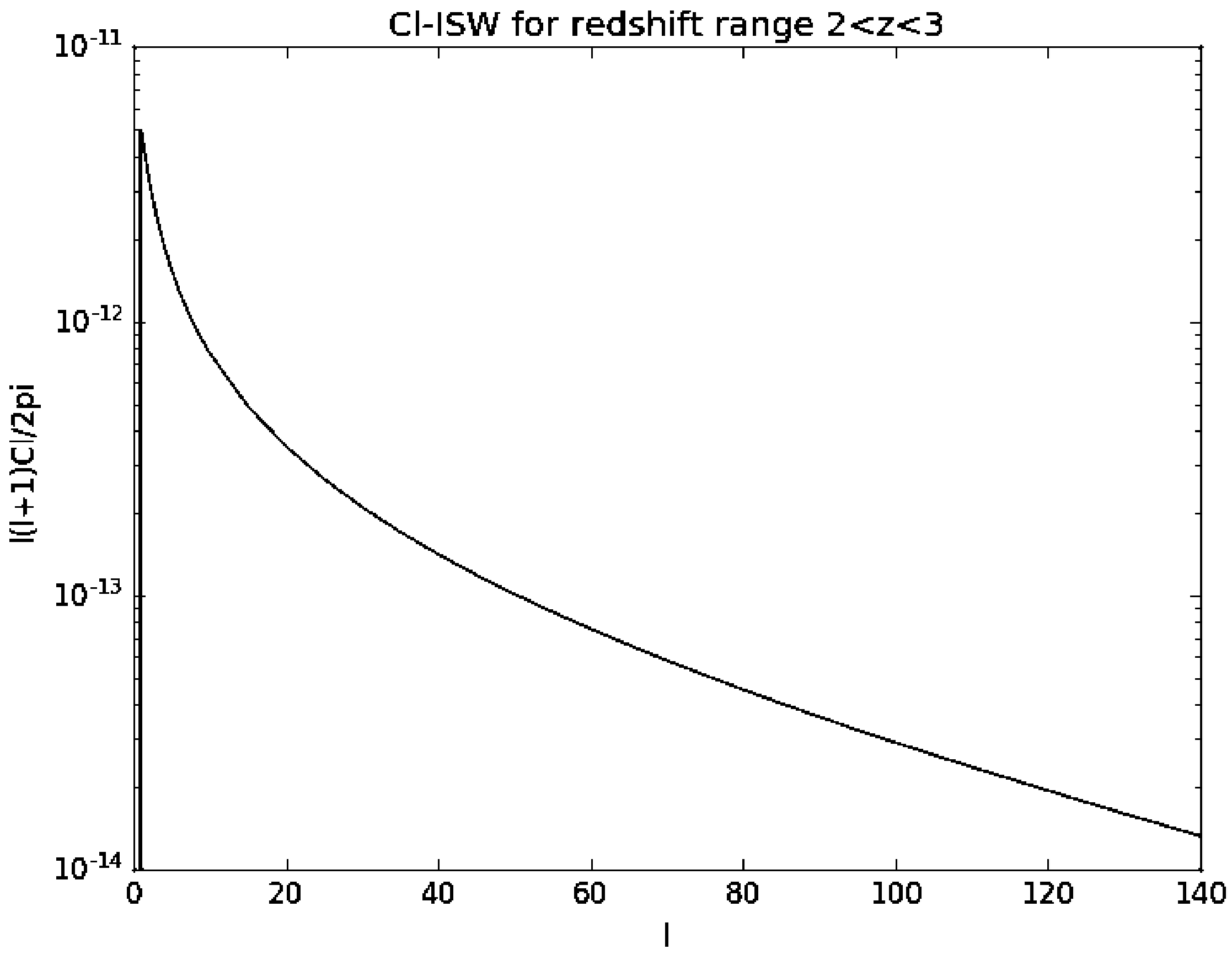

Figure 5- Theoretical ISW angular power spectrum between redshifts 2 and 3 using WMAP 9 years parameters.

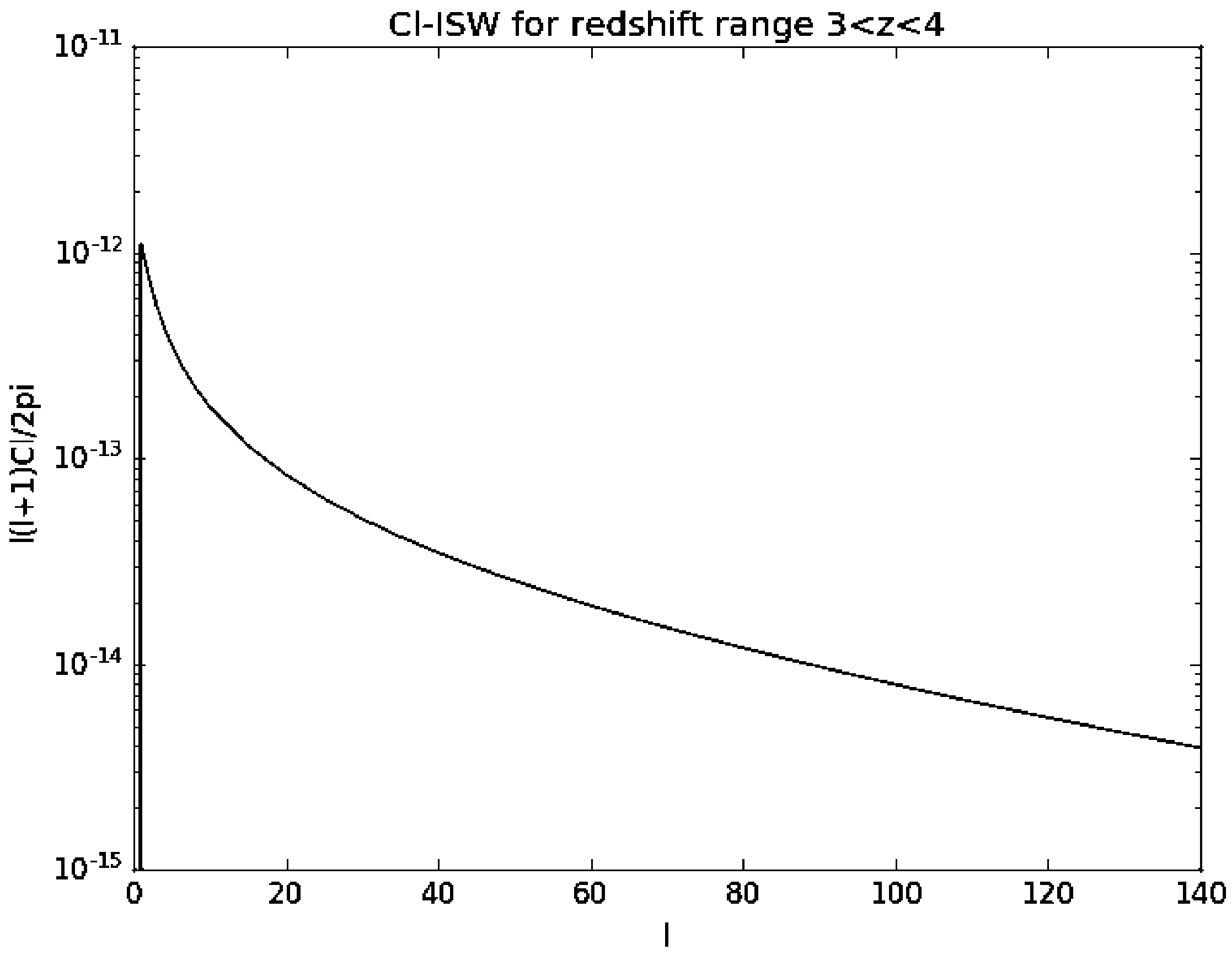

Figure 6- Theoretical ISW angular power spectrum between redshifts 3 and 4 using WMAP 9 years parameters.

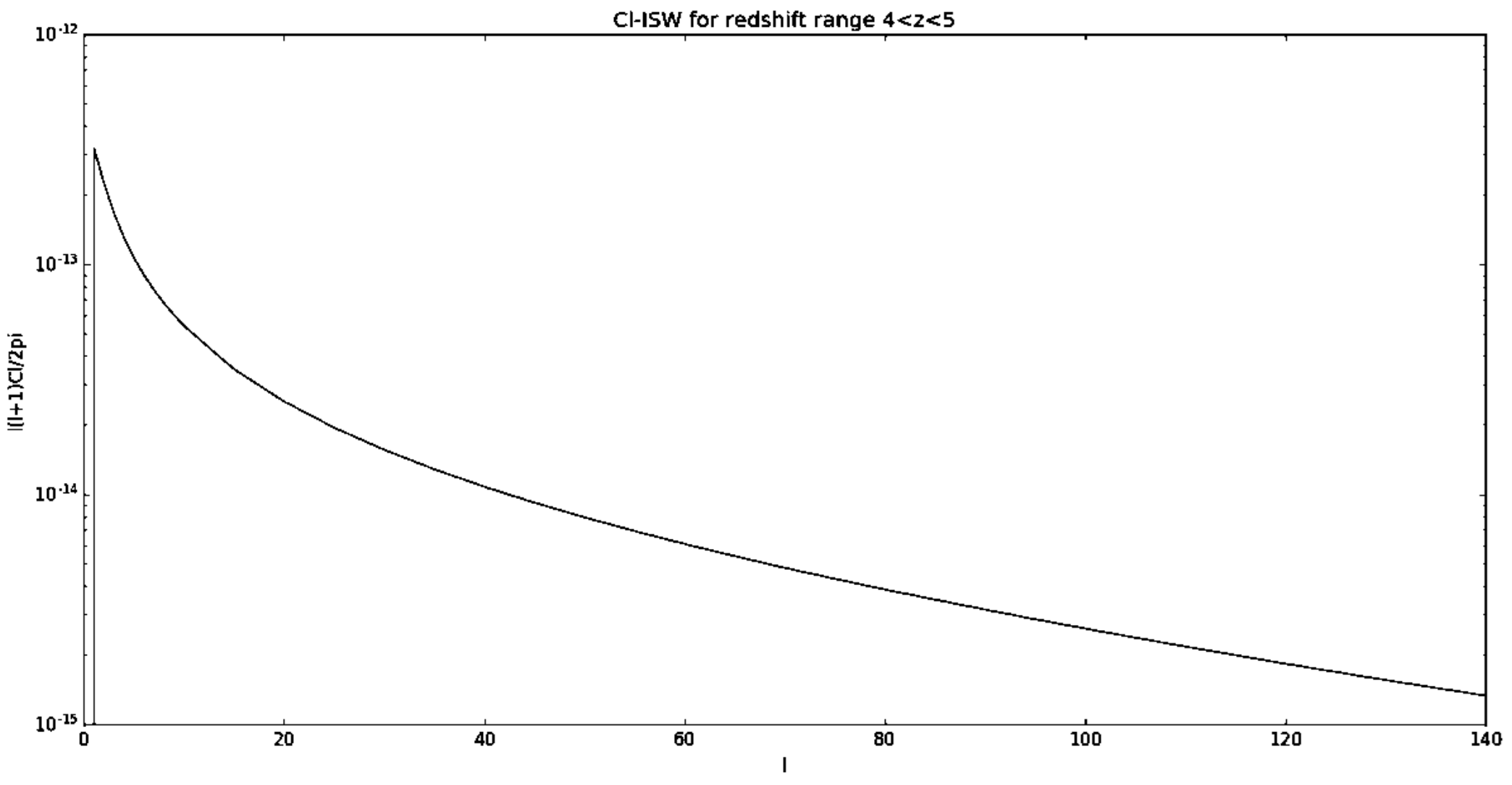

**Figure 7- Theoretical ISW angular power spectrum between redshifts 4 and 5 using WMAP 9 years parameters.**

Cosmic microwave background maps may not be able to give us the late time integrated Sachs-Wolfe effect signal directly but using large scale all sky surveys [5,22,45], the late time integrated Sachs-Wolfe effect can be measured by cross-correlating the large scale structure over/under-density maps with the cosmic microwave background anisotropy maps. We can observe from figures (3, 4, 5, 6 & 7) the relatively small ISW signal from redshift ranges beyond 2. We can also check this behavior by using the cmb-galaxy over/under-density angular cross-correlation power spectrum.

Theoretically, this cross-correlation angular power spectrum can be calculated as [5, 36, 46, 47, 48, 49, and 50]:

$$Cl^{gt}=\langle al_{mg},al_{mT}\rangle = 4\pi \int_{kmin}^{kmax} \frac{dk}{k} \Delta^2(k) Wl^g(k) Wl^t(k) \qquad (25)$$

'al' values are key measures of over/under-density in a field based on the given grid or map configuration. In a matter dominated universe Clgt values will be zero as there will be no late time ISW effect and so no angular-cross correlation between galaxy over/under-densities and CMB anisotropies. Cross-Correlation function between CMB and LSS can be expressed as:

$$w_{gT}(\theta) = \sum_{l=0}^{\infty} \frac{2l+1}{4\pi}\, Cl^{gt}\, Pl(cos\theta)$$

In our theoretical analysis, we used simulated redshift counts from S-cubed simulated database of extra galactic radio continuum ($S^3$- SEX) for square kilometer array design studies (SKADS) [51]. We used EMU-ASKAP [22, 5] 5-sigma sources parameters for our analysis.

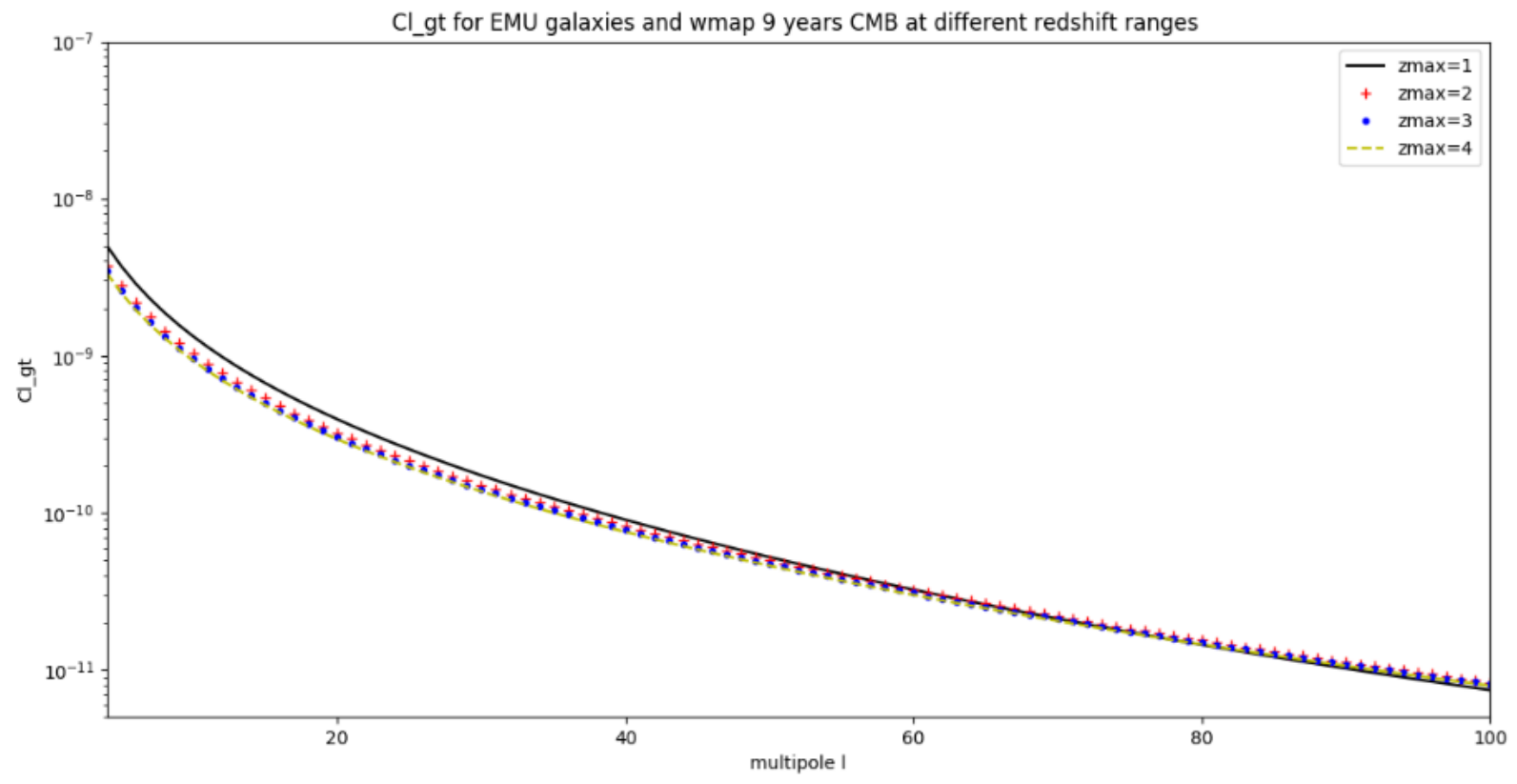

**Figure 8 - Theoretical ISW cross-correlation angular power spectrum for EMU-ASKAP galaxies over different redshift ranges using WMAP 9 years parameters. In a matter**

**dominated universe with no dark energy density, there should be no ISW cross-correlation angular power spectrum.**

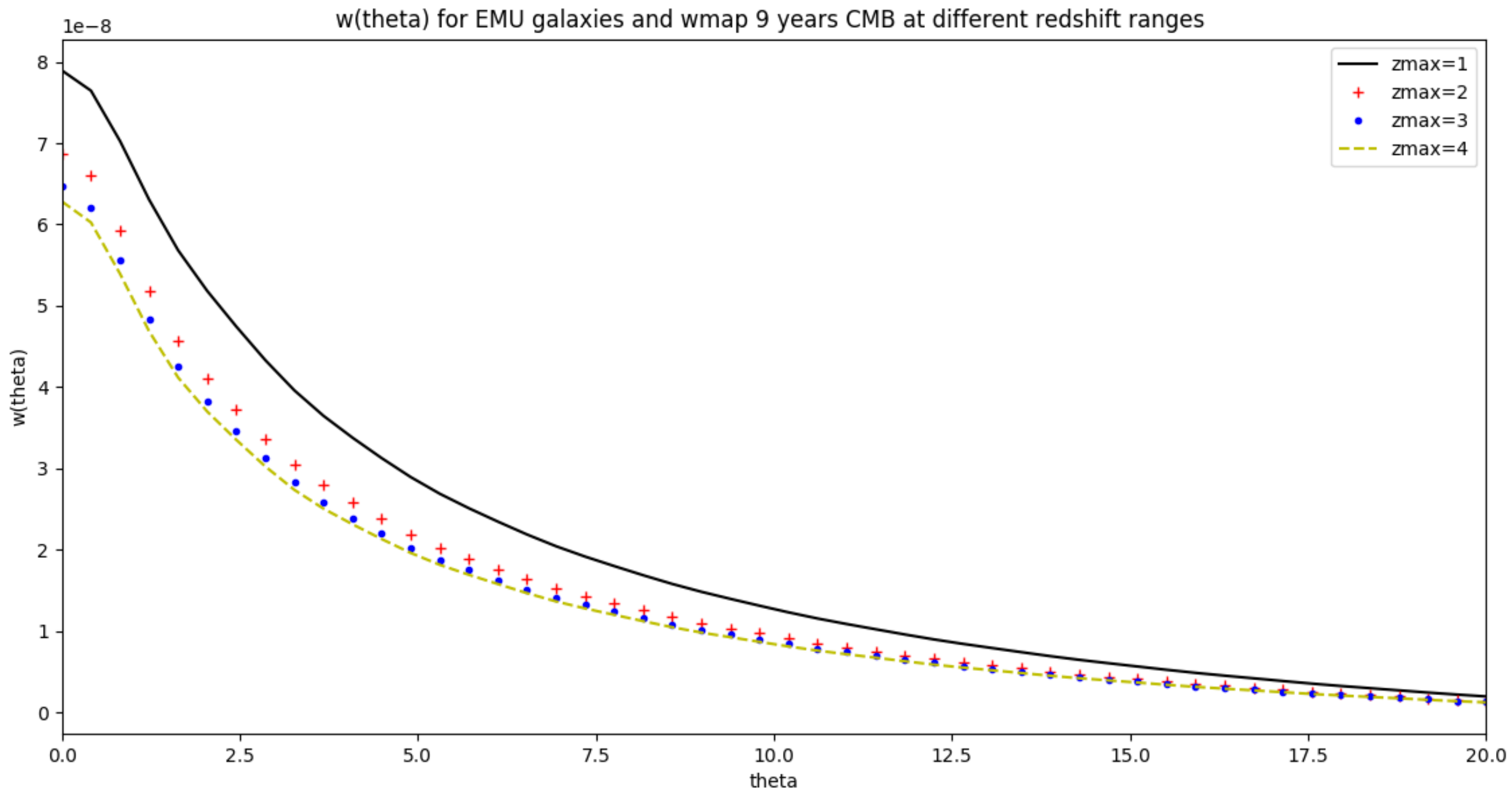

**Figure 9- Theoretical ISW cross-correlation function values for EMU-ASKAP galaxies over different redshift ranges using WMAP 9 years parameters.**

Here is $\Delta^2(k)$the logarithmic matter power spectrum, which can be calculated as:

$$\Delta^2(\mathrm{k})=\frac{k^3}{2\pi^2}P(k) \qquad (26)$$

In this, P(k) is the matter power spectrum. We use CAMB [52] to calculate P(k). $Wl^g(k)$and $Wl^t(k)$represent galaxy and ISW window functions. Galaxy window function can be written as [41, 5]:

$$Wl^g(k) = \int dz\frac{dN}{dz}b(z)g(z)jl(k\chi(\mathrm{z})) \qquad (27)$$

A more simplified form of ISW window $Wl^t(k)$ function can be written as [41, 5] :

$$Wl^t(k) = 3Tcmb\Omega m\left(\frac{H0}{ck}\right)^2 \int dzG(z)jl(k\chi(z)) \quad (28)$$

Where, jl(x) is spherical Bessel function, $g(z)=\frac{D(z)}{D(0)}$ denotes the growth factor, $\chi(z)=c\eta(z)$ denotes the commoving distances, with $\eta(z)$ as conformal loop back time and b(z) is galaxy bias as a function of redshift. Also in ISW window function, Tcmb is the average CMB temperature, G(z) is calculated as d[D(z)(1+z)/D(0)]/dz , c is the speed of light and H0 is the Hubble constant.

For large scales, linear growth factor can be written approximated as [5, 53]:

$$D(z)=\frac{5\Omega m(z)}{2(1+z)}\left\{\Omega_m(z) - \Omega_\Lambda(z) + \left[1 + \frac{\Omega m(z)}{2}\right].\left[1 + \frac{\Omega\Lambda(z)}{70}\right]\right\}^{-1} \quad (29)$$

Where,

$\Omega_m(z) = \Omega_m(1 + z)^3 /E(z)^2$

$\Omega_\Lambda(z) = \Omega_\Lambda/E(z)^2$

Here, $E(z)=\sqrt{\Omega_\Lambda I(z)+\Omega r(1+z)^4+\Omega m(1+z)^3}$ for flat Lambda-CDM model. I(z) depends on the parametrization of the dark energy EoS and for standard Lambda-CDM model with Equation of State w(z)=-1 (constant), I(z) becomes ‘1’ .

In figure (8), we can see the $Cl^{gt}$ estimates for ISW effect based on EMU-ASKAP survey parameters. In figure (9), we can observe the effect through the cross-correlation function. In both figures, there is a significant difference between zmax 1 & 2 but not much value is added after zmax=2. In order to measure the cosmological parameters, we need to use χ^2 as used in equation (7). Equation (7) used the distance modulus as it was to measure the cosmological parameters using the distance modulus obtained from the Type Ia supernovae but here for the late time ISW effect, we can constrain the parameters using the theoretical '$Cl^{gt}$' values, observed '$Cl^{gt}$' values and observed standard deviations in the '$Cl^{gt}$'. For our initial testing, we use the flat Lambda-CDM model with a constant equation of state parameter by setting w(z)=-1.

**<u>SNR calculations:</u>**

Signal to noise ratio calculations can be performed using the relation:

$$SNR = \sqrt{\sum_l} \left(\frac{Cl^{gt}}{\sigma Cl^{gt}}\right)^2$$

Error on each multipole of the angular power spectrum can be calculated as [41, 5]:

$$\sigma\, ^2{}_{CgT} = \frac{1}{fsky(2l+1)}\{ (Cl^{gt})^2 + Cl^{tt}[Cl^{gg} + 1/N]\}$$

with $Cl^{tt}$ and $Cl^{gg}$ being auto correlations of temperature anisotropy and galaxy field angular power spectrums respectively, N is the mean number of sources per steradian and $f_{sky}$ is the fraction of sky covered.

Here, $Cl^{gg}$ can be calculated as:

$$Cl^{gg} = 4\pi \int_{kmin}^{kmax} \frac{dk}{k} \Delta^2(k)\{Wl^g(k)\}^2 \qquad (30)$$

We used limber approximation over large scales [42, 43] to calculate $Cl^{gg}$ . In figure (10), we can see $Cl^{gg}$ estimates for EMU-ASKAP over zmax ranges 1,2,3 and 4.

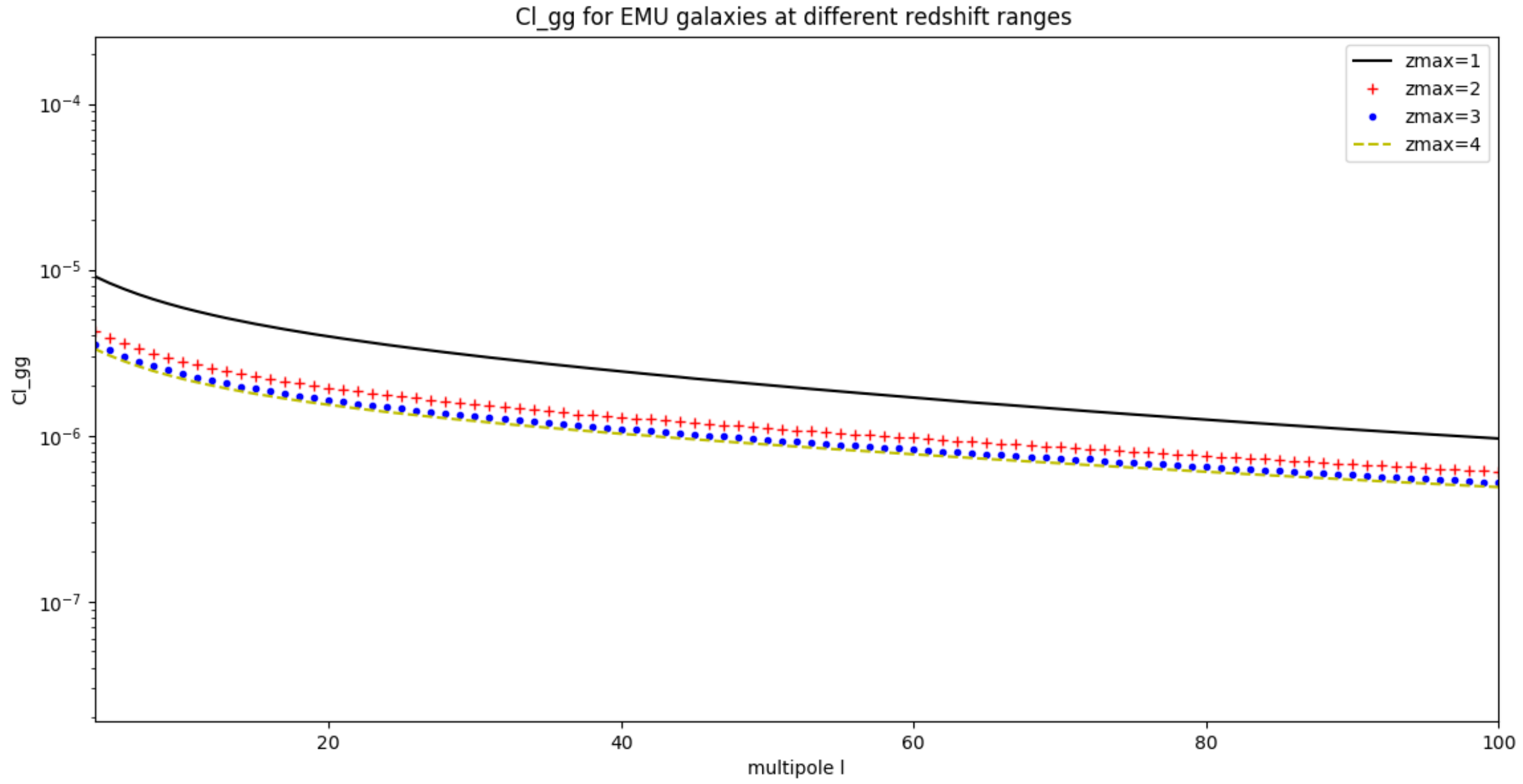

**Figure 10- Theoretical galaxy auto-correlation angular power spectrum for EMU-ASKAP galaxies over different redshift ranges using WMAP 9 years parameters.**

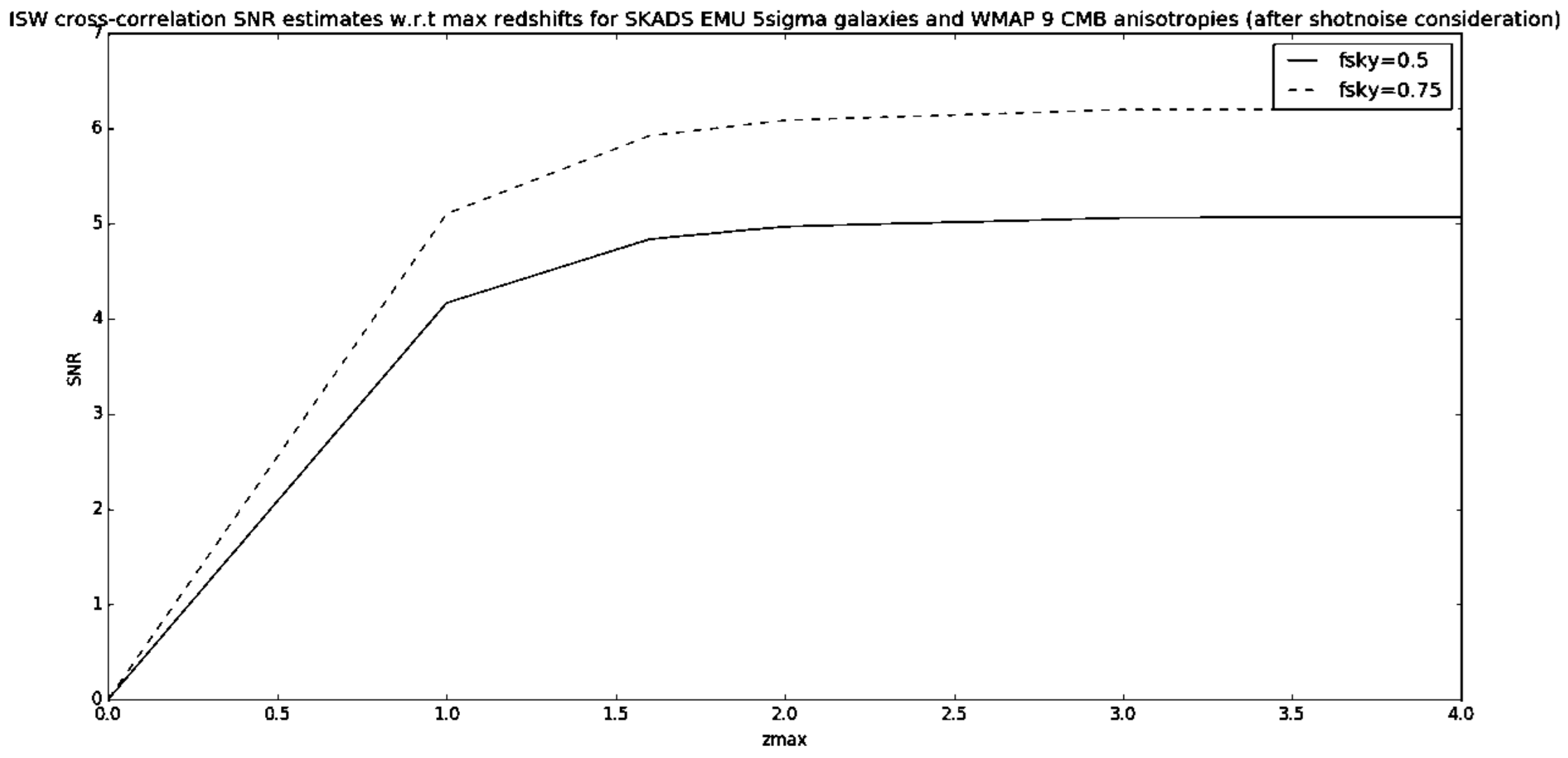

**Figure 11-Theoretical SNR ratios for ISW effect using EMU-ASKAP. Apart from dependency on redshift, shot-noise and lmax, SNRs are also dependent on the sky coverate ratio, fsky. EMU-**

**ASKAP combined with similar northern sky survey like WODAN [45] will result in improved fsky ratios and consequently improved SNRs.**

In figure (11), we can see the effects of maximum redshift range selection on the calculations of SNRs. Figure (11) shows that at zmax≥2, $Cl^{gt}$ detection through EMU-ASKAP will be ~4.5-5 for fsky=0.5 and ~5.7-6 if we use fsky=0.75 i.e combining WODAN maps after foreground removal . These SNR levels will make EMU ISW the most significant ISW results of its times [1, 7, 12, 21, 48, 49]. One thing which we can see from signal to noise ratio plots is the dependency of the error analysis and measuring the significance of our results on the galaxy auto-correlation power spectrum $Cl^{gg}$. This is also applicable to other methods such as fisher matrix [5, 54, and 55]:

$$F^{ij}=\text{fsky}\sum_{l}\frac{(2l+1)\partial Cl^{gt}}{\partial\Theta i}\, Cov^{-1}(l)\,\frac{\partial Cl^{gt}}{\partial\Theta j} \qquad (31)$$

Where, $Cov^{-1}=[Cl^{gt}]^2+(Cl^{ISW}+Nl^{cmb})(Cl^{gg}+Nl^{gg})$.

In case of the CMB measurements, we cannot directly distinguish $Cl^{ISW}$ from the whole signal and so we use the whole map predicted signal to perform the anlaysis by using $Cl^{ISW}+N^{lcmb}\approx Cl^{tt}$. Therefore, it will be useful that how $Cl^{gt}$ and $Cl^{gg}$ relationship hold up over various redshift ranges.

One thing which significantly affects the $Cl^{gg}$ measurements is the shot-noise. Shot-noise is the major source of noise in the galaxy over/under-density maps and so we can write $Nl^{gg}\approx$ Shot-noise. Shot-noise, basically accounts for the fact that we use a discrete quantity i.e.

galaxy source count as a proxy for a continuous quantity i.e. galaxy over/density. Shot-noise can simply be written as:

Shot-noise=1/Ns

Here, Ns is the average number of sources per steradian. Then the ratio between galaxy-cmb cross-correlation signal and galaxy-autocorrelation signal becomes $Cl^{gt}/(Cl^{gg}$ +Shot-noise). It is because of the shot-noise effect, along with rms confusion, that surveys like EMU-ASKAP or possible future surveys using the Square Kilometer Array (SKA)[22] can potentially provide much better quality surveys then their predecessor like the NRAO VLA Sky Survey (NVSS) [5, 56, 57, 58, 59, 60, 61, 62].

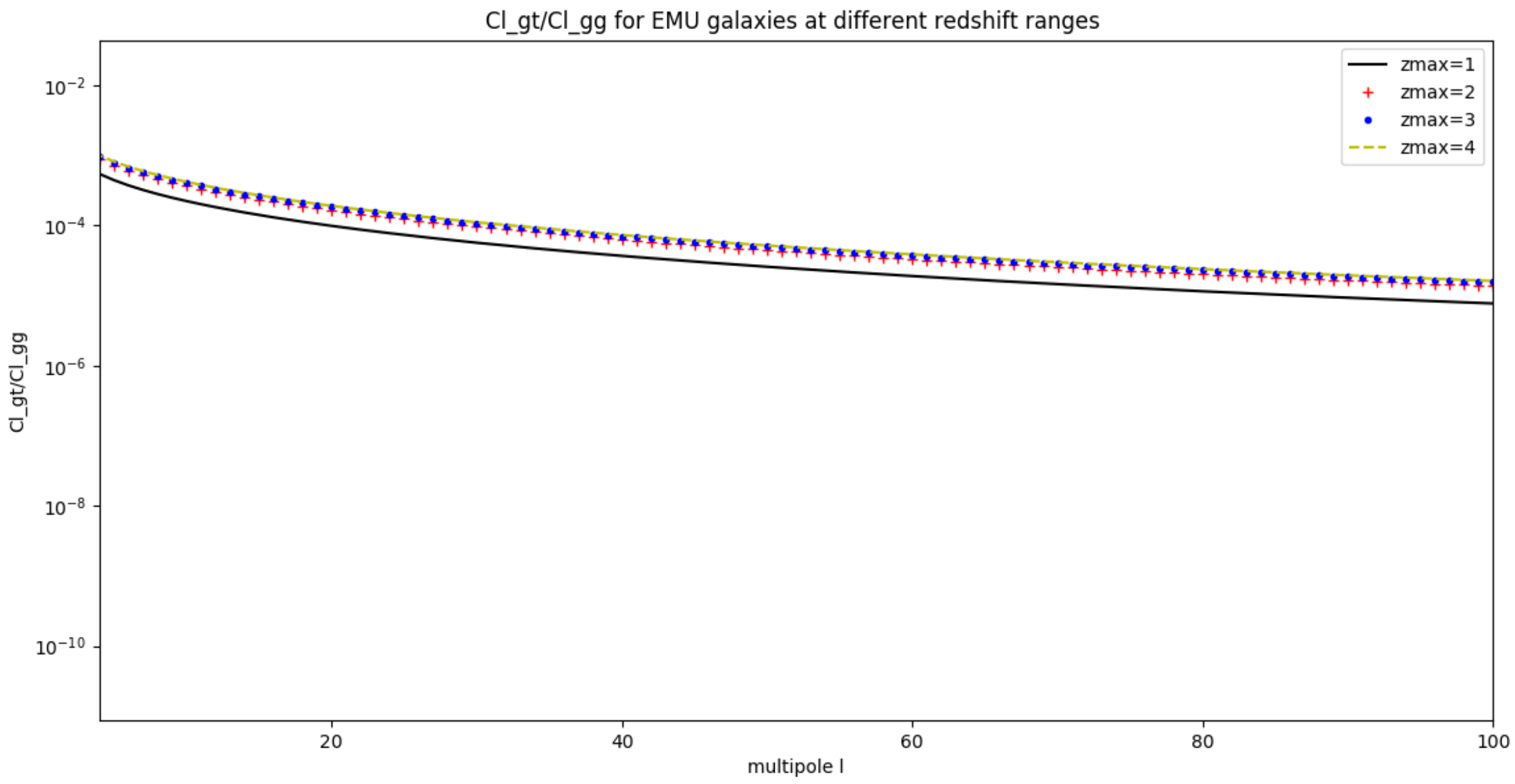

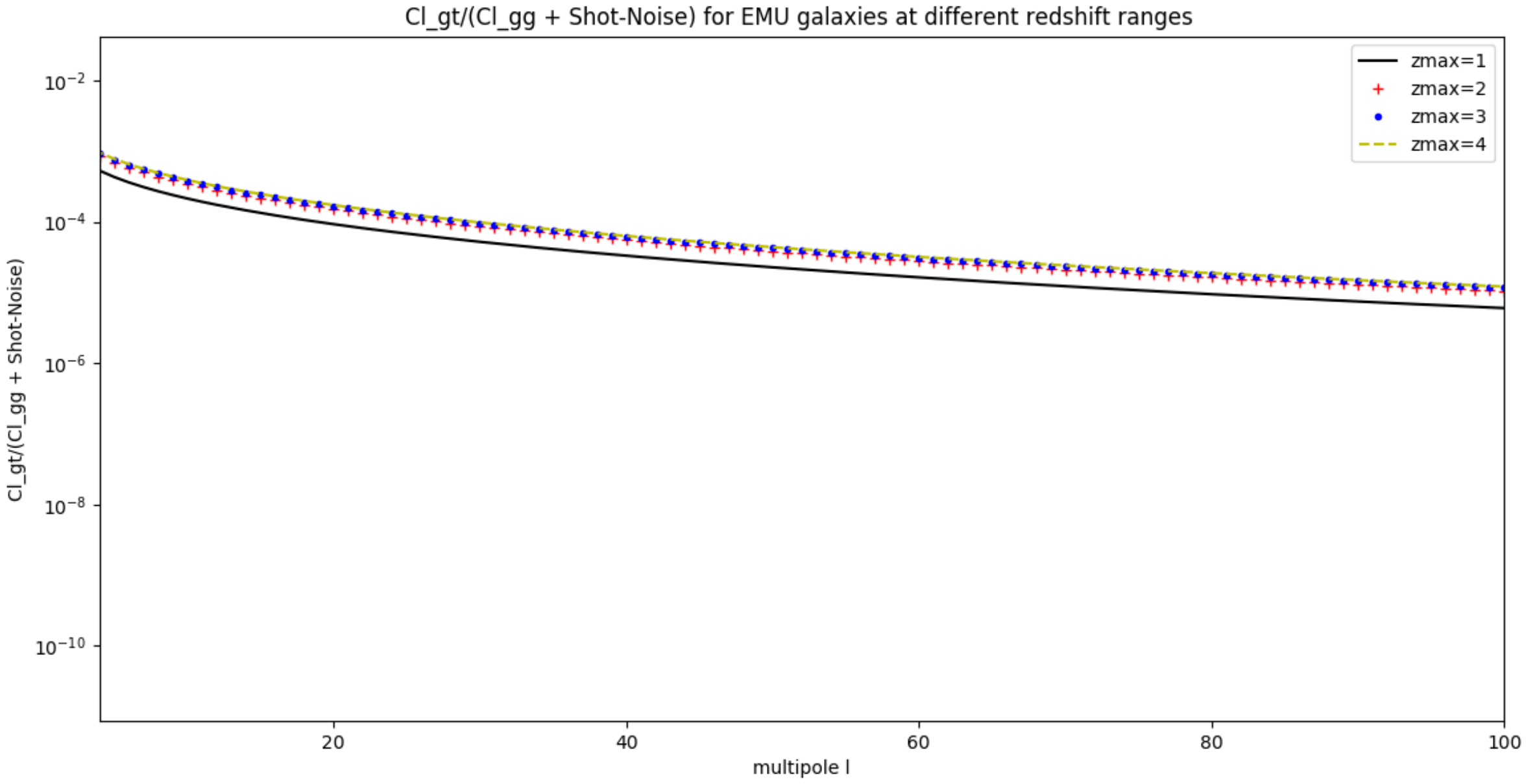

**Figure 12- Theoretical ISW cross-correlation angular power spectrum and galaxy autocorrelation ratios for EMU-ASKAP galaxies over different redshift ranges using WMAP 9 years parameters. (Upper) Without considering shot-noise. (Lower) With shot noise consideration. There is very small difference between amplitudes of the ratios but the difference is relatively more visible at higher multipole values where shot-noise plays relatively important part.**

We can see, in figure (12), how the ratio $Cl^{gt}/Cl^{gg}$ changes significantly from zmax=1 to zmax=2 but after that the affect diminishes. One reason for this behavior is the distribution of galaxies over various redshift ranges and the other reason, which relates to the accelerated expansion of the universe, is the dominance of dark energy at redshift ranges z<=1.5~2 and reducing impact in higher redshift ranges. This behavior can also be understood by the relative size of ISW signal in theoretical predictions of the CMB $Cl^{tt}$ signal as shown in figures (3,4,5,6,7). However, the overall ISW observed signal will be affected by the temperature-magnification correlations at higher redshifts which can be used to probe any signs of early dark energy signatures or to test various models of modified gravity [63].

## Estimates for the Dynamic Dark Energy Equation of State (EoS)

The late time ISW effect can also be used to estimate the extensions of the Lambda-CMD models like with the dynamic dark energy equation of state(Eos) parametrization.

This will modify the E(z) and so the $\Omega_\Lambda(z)$ in equation (29) as I(z) will not be equal to a constant value of '1'. In case of CPL parametrization with w(z) given as[64,65]:

$$w(z) = w_0 + w_a \left(\frac{z}{1+z}\right)$$

Which gives:

$$I(z) = (1+z)^{3(1+w_0+w_a)} \exp\left(-3\frac{w_a z}{1+z}\right)$$

We can see that with $w_0$=-1 and $w_a$=0,this will give a constant w(z)=-1 and so a constant I(z)=1 which are consistent with standard Lambda_CDM model we used in previous sections. To see how the late time ISW effect can be affected by the changes in EoS parameter values, we tested two set of values for our EoS with CPL parametrization with other values taken from WMAP 9 years results. First we set, $w_0$=-0.9 and $w_a$=-0.3, which gives Clgt and Clgt/Clgg ratios as given in figures (13 and 14).

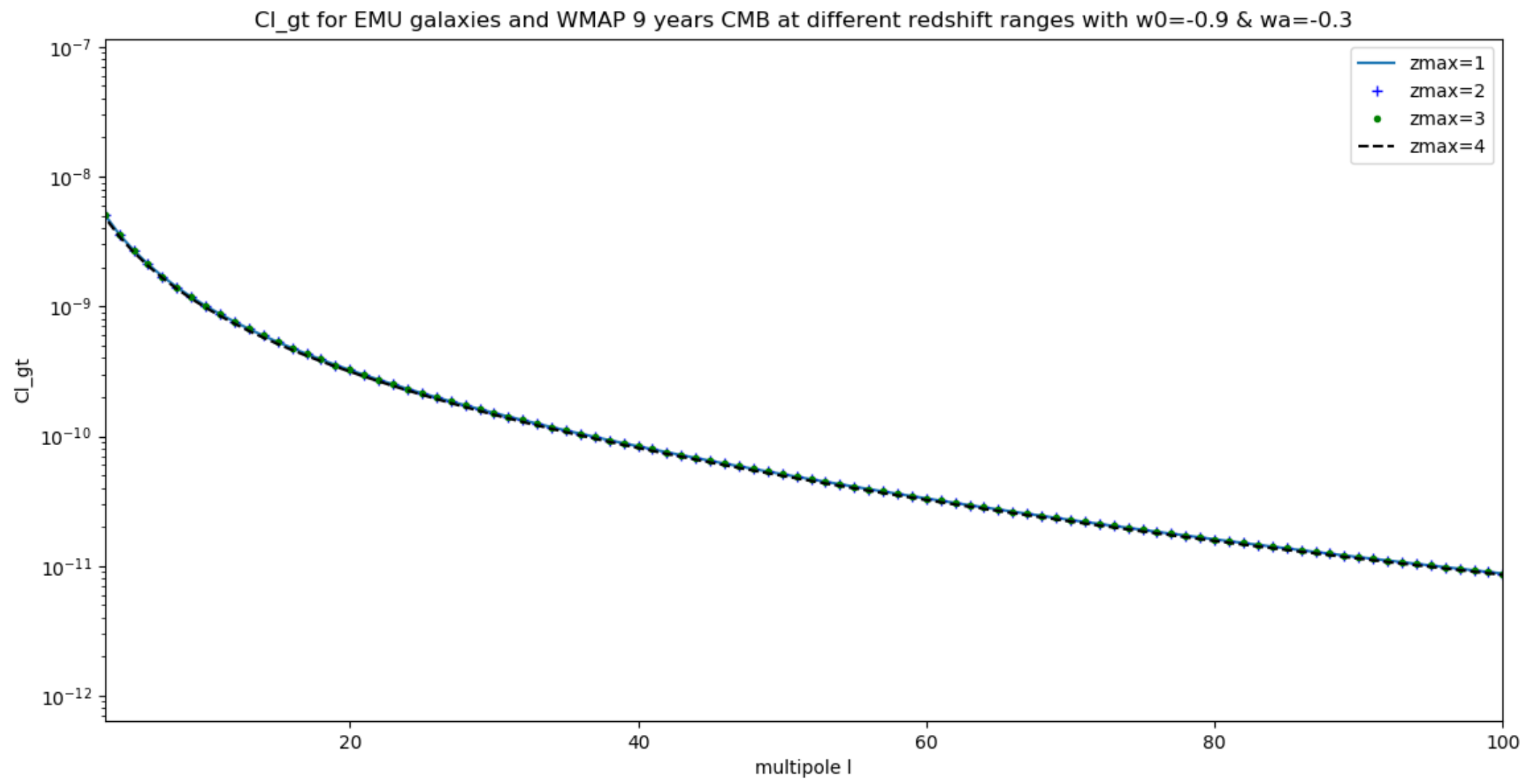

**Figure 13 - Theoretical ISW cross-correlation angular power spectrum for EMU-ASKAP galaxies over different redshift ranges using WMAP 9 years Lambda-CDM parameters with extended parameters of $w_0$=-0.9 and $w_a$=-0.3.**

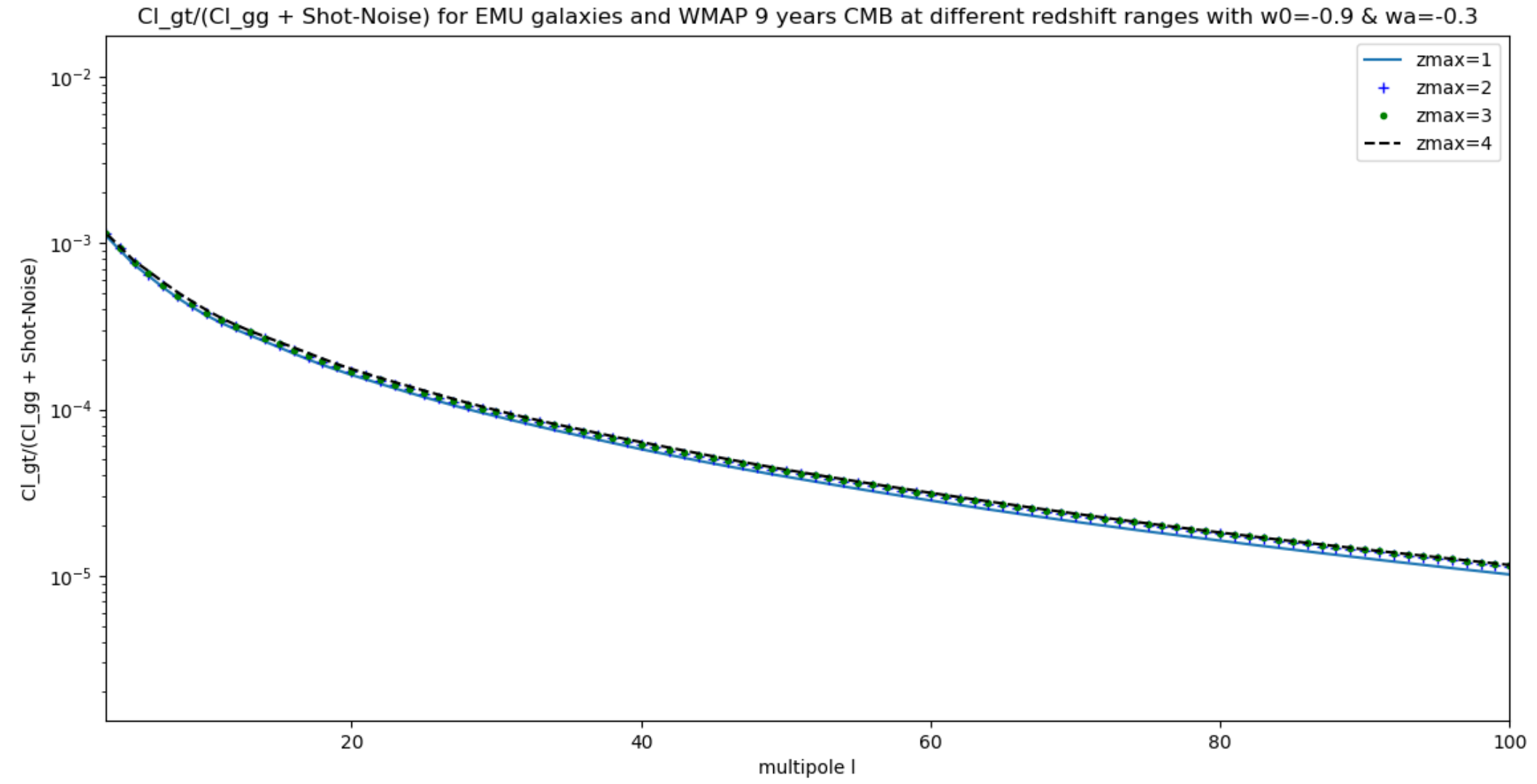

**Figure 14- Theoretical ISW cross-correlation angular power spectrum and galaxy**

**autocorrelation ratios for EMU-ASKAP galaxies over different redshift ranges using WMAP 9 years parameters with extended parameters of $w_0$=-0.9 and $w_a$=-0.3.**

Using the same parametrization, we changed $w_a$ from -0.3 to 0.3 to see how it affects the cross-correlations. The results are shown in figures (15 & 16).

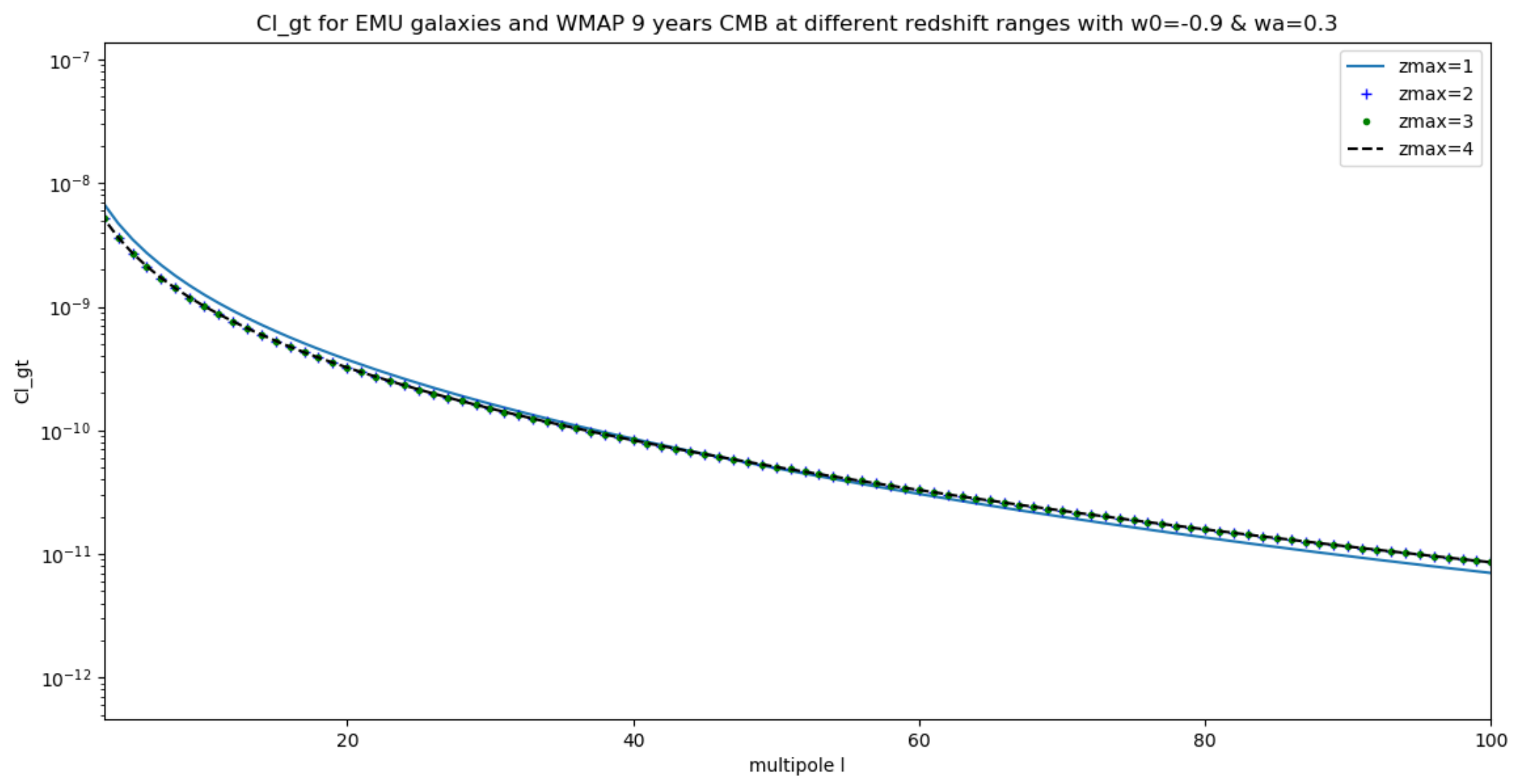

**Figure 15 - Theoretical ISW cross-correlation angular power spectrum for EMU-ASKAP galaxies over different redshift ranges using WMAP 9 years Lambda-CDM parameters with extended parameters of $w_0$=-0.9 and $w_a$=0.3.**

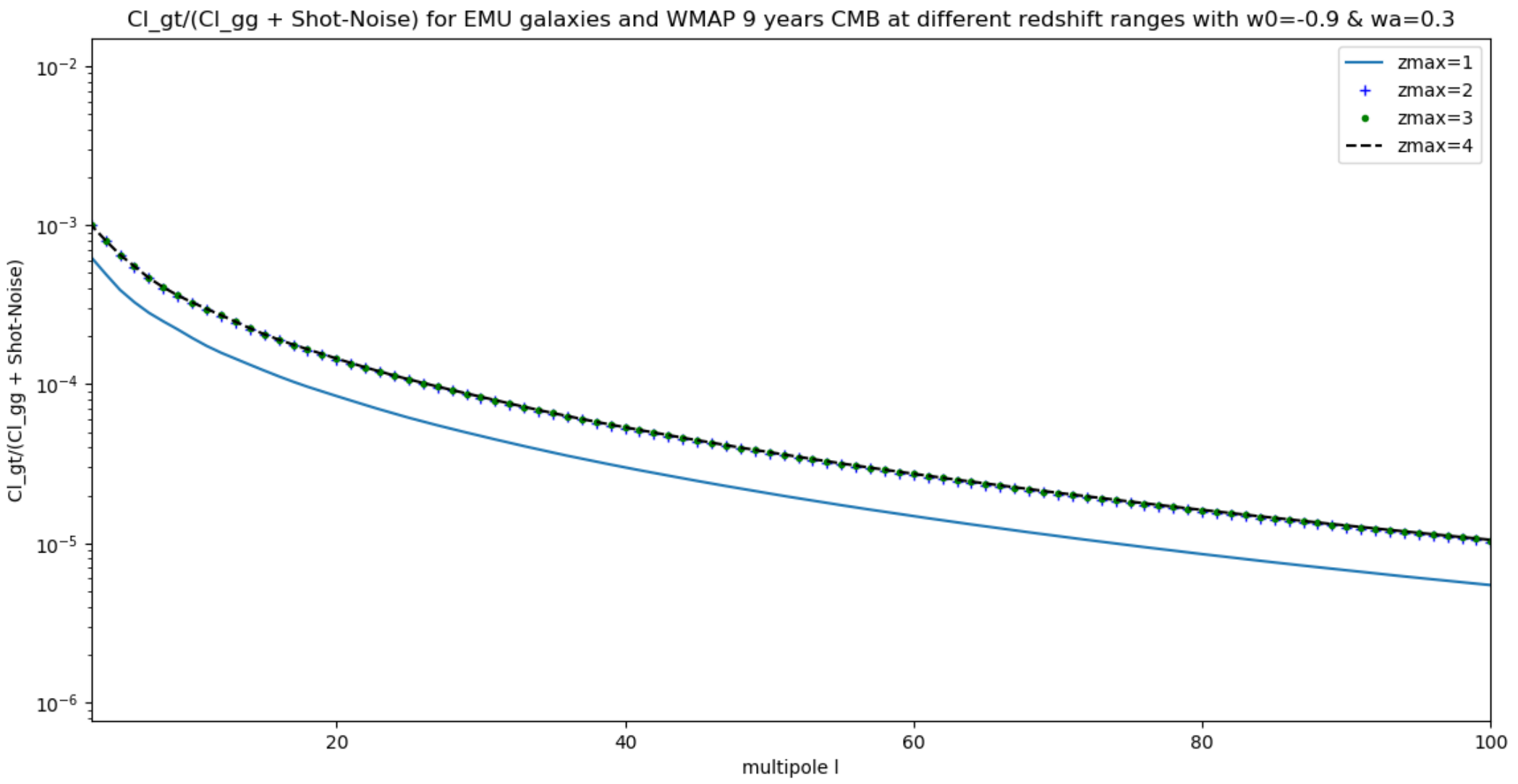

**Figure 16- Theoretical ISW cross-correlation angular power spectrum and galaxy autocorrelation ratios for EMU-ASKAP galaxies over different redshift ranges using WMAP 9 years parameters with extended parameters of $w_0$=-0.9 and $w_a$=0.3.**

We can see that the results in figures (13, 14, 15 & 16) are quite different from figures (8 & 12) which means EoS parameters can potentially have a significant role in measuring the ISW effect and so the ISW effect can be a useful tool in constraining the EoS parameters themselves. However, constraining EoS parameters will require redshift information of the galaxies in the survey.

Therefore, it will be important to estimate statistical redshifts as only about 2% of EMU sources are likely to have known redshifts [66]. This means we will need to rely on redshift bin estimates or probabilistic models to get dN/dz of our data and then match the resulting theoretical Cls or w(theta)s with the observed Cls and w(theta)s to constraint cosmological parameters. We can further improve our chances of obtaining some good estimates using modern data science and machine learning tools as analyzing ~70 million galaxies or even their sub-samples will be a serious challenge [66].

The nature of ISW effect and tracing it through galaxy surveys and CMB cross-correlations will likely continue to be a topic of high importance. One major area of importance is the study of CMB cold spot [67, 68]. Recent 2dF-VST ATLAS Cold Spot galaxy redshift survey [67] studied 7000 galaxies in the CMB 10 degrees wide cold spot core and concluded that ISW effect doesn't provide sufficient evidence for the CMB cold spot. This makes the role of futuristic surveys like SKA or EMU-ASKAP (along with advanced optical surveys) more important in understanding the phenomenon.

## Conclusion:

In this study, we first provided an overview of the accelerated expansion which was discovered using the Type 1a supernovae and how this accelerated expansion phenomenon affect the distribution of large scale structures in our universe. We then discussed the late time integrated Sachs-Wolfe effect in which the photons from the cosmic microwave background radiation are blue-shifted or red-shifted -due to the presence of large scale structures of super-voids. We also discussed about the difficulty in detecting the late time Integrated Sachs-Wolfe effect through direct measurements and how galaxy surveys like EMU-ASKAP can help in observing the effect through cross-correlation angular power spectrum analysis. We also discussed some estimates over different redshift ranges for ISW detection through the upcoming EMU-ASKAP survey. We also observed the diminishing benefit of increasing the redshift ranges, due to the galaxy distribution and lack of dark energy density effects over larger redshift ranges in the Lambda-CDM, in the ISW studies using theoretical SNR analysis and the ratio of $Cl^{gt}/Cl^{gg}$. However, we also observed that the late time ISW effect will behave differently under the extended Lambda-CDM models like using the dynamic dark energy EoS models which will also make it highly useful to obtain the observed or statistical redshift information for the galaxies in the survey.